\documentclass{article}

\usepackage{amssymb}
\usepackage{amsmath}
\usepackage{amsfonts}
\usepackage{rotating}
\usepackage{graphicx}
\usepackage[utf8]{inputenc}
\usepackage[T1]{fontenc}
\usepackage{lmodern}
\usepackage{graphicx}
\usepackage{array}
\usepackage{natbib}
\usepackage{bm}
\begin{document}

\title{\vspace{-10mm}Estimating the treatment effect on the treated under
  time-dependent confounding in an application to the Swiss HIV Cohort Study} \date{} 

\author{Jon Michael Gran${}^{1,*}$ \and Rune Hoff${}^{\text{2}}$ \and Kjetil
  R{\o}ysland${}^{\text{2}}$ \and Bruno
  Ledergerber${}^{3}$ \and James Young${}^{4}$ \and Odd
  O. Aalen${}^{2}$}

\maketitle

{\footnotesize \emph{${}^{1}$Oslo Centre for Biostatistics and Epidemiology,
Oslo University Hospital and University of Oslo, Norway.} }
{\footnotesize \emph{${}^{2}$Oslo Centre for Biostatistics and Epidemiology,
University of Oslo, Norway.} }
{\footnotesize \emph{${}^{3}$Division of Infectious Diseases and Hospital
Epidemiology, University Hospital Zurich, University of Zurich, Switzerland.}
}
{\footnotesize \emph{${}^{4}$Basel Institute for Clinical Epidemiology and
Biostatistics, University Hospital Basel, Switzerland.} }
{\footnotesize \emph{${}^{\ast}$Address for correspondence: Jon Michael Gran, Oslo
Centre for Biostatistics and Epidemiology, Department of Biostatistics,
University of Oslo, P.O.Box 1122 Blindern, 0317, NORWAY. E-mail: j.m.gran@medisin.uio.no.}}

\begin{abstract}
  When comparing time-varying treatments in a non-randomised setting,
  one must often correct for time-dependent confounders that influence
  treatment choice over time and that are themselves influenced by
  treatment. We present a new two step procedure, based on additive
  hazard regression and linear increments models, for handling such
  confounding when estimating average treatment effects on the treated
  (ATT). The approach can also be used for mediation analysis. The
  method is applied to data from the Swiss HIV Cohort Study,
  estimating the effect of antiretroviral treatment on time to AIDS or
  death. Compared to other methods for estimating the ATT, the
  proposed method is easy to implement using available software
  packages in R.
  
\end{abstract}

\section{Introduction}

The issue of time-dependent confounding is central in causal
inference. When comparing time-varying treatments in a non-randomised
setting, one will often need to correct for confounders that influence
the treatment choice over time, and that are themselves influenced by
treatment. Such confounding can be present in both clinical and
epidemiological data, and requires a careful analysis. The inverse
probability weighted marginal structural Cox model has become the most
popular tool for handling such confounding in settings with
time-to-event outcomes \citep{robins00, sterne05}. The method
identifies a marginal estimate, where the effects of the
time-dependent confounders on treatment are removed by inverse
probability weighting procedures. Two other general methods for
handling time-dependent confounding has also been developed;
g-computation \citep{keil2014, westreich2012, cole2013, edwards2014,
  taubman2009} and g-estimation \citep{vansteelandt14b,
  picciotto16}. All these three approaches fall under the so called
g-methods by Robins and co-authors \citep{robins2009}; see
e.g. \citet{daniel13} for an introduction to these methods. Other ways
of dealing with time-dependent confounding has also been suggested,
such as the sequential Cox regression approach \citep{gran10}.

Treatment effects may be viewed in various ways and there are
typically more than one unique causal estimand that can be
defined. The most common methods used to adjust for time-dependent
confounding, inverse probability weighting and g-computation, are
typically used to estimate the average total treatment effect
(ATE). Another causal contrast is the average treatment effect on the
treated (ATT). The ATT can in principle also be identified using
existing methods, and especially through g-estimation of structural
nested models, but this is hardly done in practice \citep{li14,
  vansteelandt14b}. The reason is partly that it is not a
straightforward method to implement, and, when working with
time-to-event outcomes, g-estimation is typically developed using
accelerated failure time models and not in a more traditional hazard
regression framework.

In this paper we present a new method for estimating the ATT, which is
based on combining two off-the-shelf methods; hazard regression models
and a method for modelling missing longitudinal covariate
trajectories. We will discuss the differences between the ATE and ATT
effect measures for time-varying treatments, when the latter can be of
greater interest, and compare results from methods estimating both
quantities. The idea behind our proposed method is to estimate the
values that possible confounding processes would have had if,
counterfactually, a treated person had not been treated. If these
counterfactual values are substituted for the actual observed values,
then, given some assumptions, one can estimate the ATT. In other
words, the causal effect of treatment is estimated by modelling the
treatment-free development of time-varying covariates. We shall also
show that our approach has an interesting relationship to mediation,
where quantities similar to natural direct and indirect effects can be
estimated from the analysis.

Note that the estimation of counterfactual values under the assumption
of no treatment has previously been applied in the two-stage approach
of \citet{kennedy2010} and \citet{taylor2014}, although in a different
setting than here. There is also a relationship to the prognostic
index discussed in \citet{hansen2008}.

The proposed method is developed for additive hazards regression
models \citep{aalen80,aalen89}. The method is applied to data from the
Swiss HIV Cohort Study, analysing the ATT of antiretroviral treatment
on time to AIDS or death for HIV patients. The method is also explored
in a accompanying simulation study. Incidentally, simulating data with
time-dependent confounding in the setting of additive hazards models
it is quite simple, as opposed to using Cox proportional hazards
models \citep{havercroft12}. The additive hazards model has generally
been shown to be useful in causal settings because it allows explicit
derivations that are not available for the Cox model
\citep{martinussen11, martinussen13, vansteelandt14, tchetgen14,
  strohmaier15}. This is due to collapsibility and other
properties. The additive model also allows a more explicit process
approach which is important in causal inference \citep{aalen14}. See
\citet{martinussen16} for a recent application of instrumental
variables and the additive hazars model for estimating the ATT.



As with the popular inverse probability weighting approach, the
suggested method can also be said to be founded on the close
relationship between the issues of missing data and causal inference
\citep{howe15}. When a treatment is initiated, data that would have
counterfactually been observed under no treatment can be considered as
missing. If these data were known, the causal effect could be easily
estimated. In this paper we use Farewell's linear increments model
\citep{farewell06, diggle07} to estimate such missing data, or more
precisely, to estimate the missing covariate trajectories of the
counterfactual time-varying variables. When counterfactual covariate
trajectories are estimated, we will show that these can be used to
give us an estimate of the ATT.

The proposed method has the advantage of focusing specifically on
individuals actually on treatment and estimating what they gain from
it, both in terms of the main outcome (survival) and intermediate
time-dependent covariate trajectories. In clinical practice the
decision to set a patient on a treatment will depend on specific
criteria regarding whether the treatment is of use for the patient.
When judging the effect of such a treatment it is natural to take this
into consideration, as is done when assessing the treatment effect for
the actually treated patients. Such an analysis clearly is a useful
supplement to the average treatment effect estimated in marginal
structural models.

\begin{figure}[ptb]
\begin{center}
\includegraphics[scale=1.15]{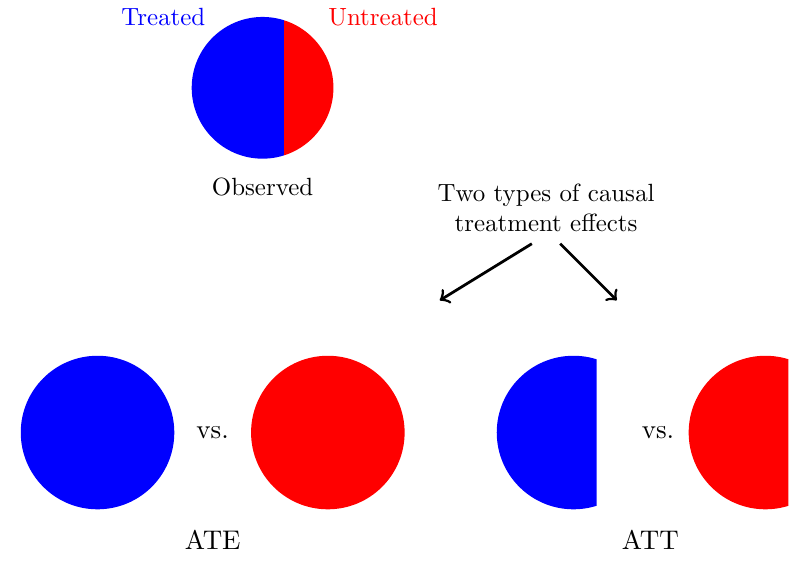}
\end{center}
\caption{Illustration of the actual observed population, compared to
  the counterfactual target populations when estimating average
  treatment effects on the treated (ATT) and average treatment effects
  (ATE).}
\label{ATT}
\end{figure}

The basic idea behind the ATT is illustrated in Figure \ref{ATT}. The
lower figures on the left hand side illustrate a comparison between
the two treatments applied to the whole population; the average
treatment effect (ATE). The lower right hand figures give a similar
comparison, but limited to the subgroup actually on treatment; the
average treatment effect on the treated (ATT). The idea and estimation
of treatment effect on the treated compared to average treatments
effects are discussed in many papers, and a good reference is that of
\citet{pirracchio13}, where they point out that the ATT and ATE
estimates may give very different results and that the choice of
method depends on the aims of the analysis. 

For time-varying treatments, which is the main concern in this paper,
the difference between the ATE and ATT becomes more complicated than
in Figure \ref{ATT}. The ATE will then correspond to the treatment
effect in a world where treatment initiation is randomized at every
time point, which again corresponds to the average treatment effect in
a counterfactual world where everyone is observed under every possible
time of treatment start (including never treated). The ATT on
the other hand, corresponds to the average effect of the treatment
regimes that were actually observed in the study population.

Our causal model for estimating the ATT is spelt out in Section
\ref{sec:additive model}, while Farewell's linear increments model is
formally introduced in Section \ref{sec:flim}.  The relationship to
mediation is demonstrated in Section \ref{sec:mediation}. In Section
\ref{sec:shortcut} we discuss a shortcut through regression that
yields an easier implementation of the method. We also briefly discuss
estimation in Cox proportional hazards models. The application to the
Swiss HIV Cohort data is found in Section \ref{sec:shcs}, while
results from simulations are summaried in Section \ref{sec:simulation}. A
discussion is given in Section \ref{sec:discussion}. Details on the
simulations and R code for carrying out the analyses, combined with a
package for Farewell's linear increments model \citep{Hoff14}, is
available in an online appendix.

\section{A counterfactual additive hazards model for the treatment
  effect on the treated}

\label{sec:additive model}

\subsection{The causal estimand of interest}

\begin{figure}[ptb]
\begin{center}
\includegraphics[width = 0.9\textwidth]{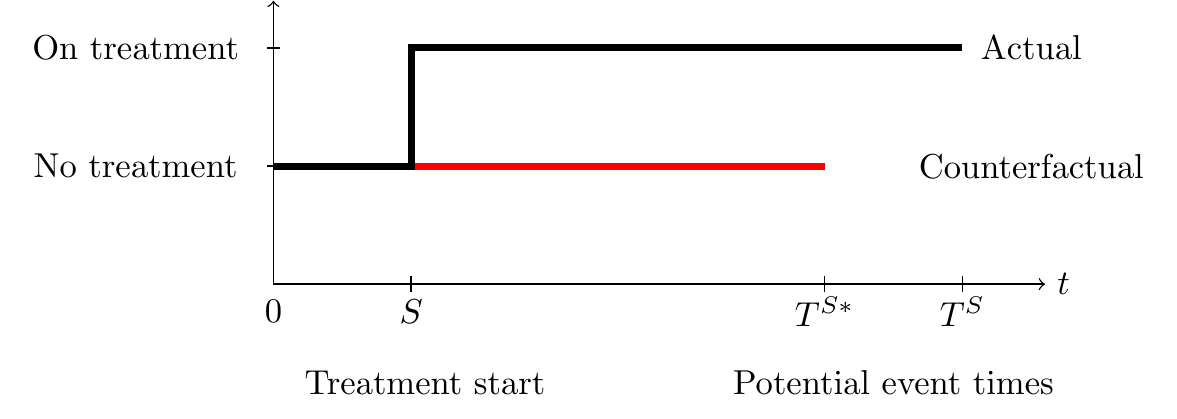}
\end{center}
\caption{ Illustration of the counterfactual scenarios in terms of
  event times, with $S$ being the time of treatment initiation, $T^S$
  the event time under the given treatment and $T^{S*}$ the event time
  given no treatment. The black line denotes the observed outcome and
  the red line the unobserved counterfactual, given that treatment was
  never initiated.}
\label{causal_setup}
\end{figure}

The target estimand in this paper is the ATT for a time-to-event
outcome in a setting with time-dependent exposure and
confounding. Consider the illustration in Figure \ref{causal_setup},
depicting the situation for a hypothetical individual that starts
treatment at some time point $S$. The time scale is time since
inclusion in the study.  When treatment is started, we can imagine a
counterfactual scenario where treatment was not started at time $S$,
and the individual remained untreated. Let $T^{S}$ be the potentially
observed time of the outcome of interest when treatment is actually
started at time $S$ and let $T^{S\ast}$ be the potential event time in
the counterfactual situation that treatment is \emph{not} started at
time $S$ (or later).

In order to achieve notational clarity, we express the hazard rates in
terms of conditional probabilities. Assume $S=s$ and let $h^1(t;s)$ be
the causal hazard rate under the actual treatment initiation at
$s$, for $t\geq s$. We can then write
\begin{align}
h^{1}(t;s)  &  =\frac{1}{dt}P(T^{S}\leq t+dt\,|\,S=s,T^{S}>t)\nonumber
\\
&  =E\left[  \frac{1}{dt}P(T^{S}\leq t+dt\,|\,S=s,L^{1}%
,T^{S}>t)\,|\,S=s,T^{S}>t\right],
\label{eq:h1}
\end{align}
where the second line in the formula follows from the Innovation
theorem \citep{aalen08}. Note that $S=s$ represents conditioning,
while $T^{s}$ is the result of a $do$-operator; that is the survival
time under $do(\mathrm{treatment}$ \textrm{starts at }$s)$, where $S$
is a random variable. $L^{1}$ is the value of the corresponding
covariate process under this same intervention.

Similarly, the causal hazard rate for the counterfactual scenario
where treatment is not started at $s$ (or later),
$h^0(t;s)$, can be written:
\begin{align}
h^{0}(t;s)  &  =\frac{1}{dt}P(T^{S\ast}\leq t+dt\,|\,S=s,T^{S\ast}>t)
\nonumber \\
&  =E\left[  \frac{1}{dt}P(T^{S\ast}\leq t+dt\,|\,S=s,L^{0},T^{S\ast
}>t)\,|\,S=s,T^{S\ast}>t\right],
\label{eq:h0}
\end{align}
where  $T^{S\ast}$ is the potential survival time
under the intervention $do(\mathrm{no}$\textrm{\
  treatment}$)$. Notice that we still condition with respect to
treatment start at time $S=s$, but that counterfactually treatment is
not started. The notation $L^{0} $\ denotes the value of the
covariate process in this case. The classic distinction in causal
inference between conditioning and intervention is important here. We
condition on treatment start at $S$, that is, we limit the outcomes to
the part of the sample space with this starting time. However, we then
intervene to insure that treatment is not really started.

The causal hazard difference is then given as follows for $t\geq s$:
\begin{align}
d(t;s)  &  =h^{1}(t;s)-h^{0}(t;s). 
\label{eq:hazdiff1}
\end{align}

Defining the causal effect of interest, we integrate the causal hazard
difference over $s$ to get a marginal effect. The effect will then
only be a function of $t$, and can be estimated. The average causal
hazard difference is thus:
\begin{align}
d^*(t)=E(d(t,S)\,|\,t>S)=E^{S}(d(t,S)),
\label{eq:targetparameter1}
\end{align}
which corresponds to the ATT for all observed versions of treatment,
started at any time point $S$. $E^{S}(.)$ denotes the conditional
expectation given $t>S$, computed over the distribution of $S$.

Note that, conceptually, this means that when treatment is started
for an individual, we want to create a "copy" of that individual who
does not start treatment and observe her or him under both
scenarios. When looking at the entire study population, this
corresponds to a randomised study at treatment start where
randomisation determines whether the intended treatment is actually
started or not. Hence, our target parameter $d^*(t)$ is the average
effect of the treatment that was actually started, compared to no
treatment.

\subsection{Assumptions}

We assume an additive hazards regression model for the event of
interest, taking the form
\begin{equation}
  \alpha(t;B,\mathbf{L})=\delta(t)B(t)+\gamma_{0}(t)
  c+\gamma_{1}(t)L_{1}(t)+\ldots+\gamma_{k}(t)L_{k}(t), 
\label{causal model}
\end{equation}
with parameters $\delta(t)$ and $\gamma_{m}(t)$, for
$m=0,\ldots,k$. Here, $c$ denotes a vector of baseline covariates,
including a constant, and $\gamma _{0}(t)$\ is a vector of
time-dependent coefficients. The covariate processes
$\mathbf{L}(t)=(L_{1}(t),\ldots,L_{k}(t))$ are various time-dependent
quantities that may influence or be influenced by treatment. These
processes are written as individual components and not in vector form
since they are the main focus of interest; they are assumed to be
measured repeatedly over time. The process $B(t)$ indicates whether
treatment has been started; it is equal to zero with no treatment and
changes to 1 when treatment starts. It is assumed that once started,
treatment continues, so $B(t)$ cannot return to zero.

Assume that the model in Equation (\ref{causal model}) is a causal
model in the following sense: If one manipulates (intervenes on) the
processes $\mathbf{L}(t)$\ or $B(t)$, then the parameter functions
$\delta(t)$ and $\gamma_{0}%
(t),\ldots,\gamma_{k}(t)$ in Equation (\ref{causal model}) are assumed
to stay unchanged. This implies that the assumption of no unmeasured
confounding is met. For causal modelling in the setting of stochastic
processes, see \citet{Roysland12}.

Note that in our observed data, by definition, $B(S)=1$, but we also
imagine a counterfactual scenario where $B(t)$ was manipulated to be 0
for $t\geq S$. In other words, we consider the situation where the
person on treatment was not actually put on treatment and ask what
would have happened then. The covariate processes of this non-treated
scenario, where treatment is \emph{not} started at time $S=s$ or later,
are denoted
$\mathbf{L}^{0}(t;s)=(L_{1}^{0}(t;s),\ldots,L_{k}^{0}(t;s))$; these
are unobserved counterfactual quantities. We will therefore assume
that we have a model for estimating these counterfactual individual
covariate processes. The procedure we use for doing so, which is
based on Farewell's linear increments model for missing longitudinal
data, and the corresponding causal assumptions are described in detail
in Section \ref{sec:flim}. The covariate process under the scenario
that treatment is actually started at time $S=s$ are denoted
$\mathbf{L} ^{1}(t;s)=(L_{1}^{1}(t;s),\ldots,L_{k}^{1}(t;s))$ and
correspond to the actually observed processes $\mathbf{L}(t)$ for
$t\geq s$.

\subsection{Identification}

The additive structure of the model in Equation (\ref{causal model}) leads to
the following formulation for the hazard rates in Equation (\ref{eq:h0}) and
(\ref{eq:h1}):
\begin{eqnarray}
h^{1}(t;s)  &  =\delta(t)+\gamma_{0}(t)
c+E\left[  \gamma_{1}(t)L_{1}^{1}(t)+\ldots+\gamma_{k}(t)L_{k}^{1}
(t)\,|\,S=s,T^{S}>t\right] \nonumber \\
&  =\delta(t)+\gamma_{0}(t)
c+\gamma_{1}(t)\tilde{L}_{1}^{1}(t;s)+\ldots+\gamma_{k}(t)\tilde{L}_{k}%
^{1}(t;s), \label{eq:h1ts}
\end{eqnarray}
where $\tilde{L}_{i}^{1}(t;s)=E\left[  L_{i}^{1}(t;s)\,|\,S=s,T^{S}>t\right]
$, and
\begin{align*}
h^{0}(t;s)  &  =\gamma_{0}(t)
c+E\left[  \gamma_{1}(t)L_{1}^{0}(t)+\ldots+\gamma_{k}(t)L_{k}^{0}
(t)\,|\,S=s,T^{S\ast}>t\right] \\
&  =\gamma_{0}(t)
c+\gamma_{1}(t)\tilde{L}_{1}^{0}(t;s)+\ldots+\gamma_{k}(t)\tilde{L}_{k}
^{0}(t;s)
\end{align*}
where $\tilde{L}_{i}^{0}(t;s)=E\left[ L_{i}^{0}(t)\,|\,S=s,T^{S\ast
  }>t\right]$.

The corresponding causal hazard difference in Equation (\ref{eq:hazdiff1}) can
then be written as follows for $t\geq s$:
\begin{align}
d(t;s)  &  = \delta(t)+\gamma_{1}(t)(\tilde{L}_{1}^{1}(t;s)-\tilde{L}_{1}
^{0}(t;s))+\ldots+\gamma_{k}(t)(\tilde{L}_{k}^{1}(t;s)-\tilde{L}_{k}
^{0}(t;s)). \label{individual causal}
\end{align}
When defining the causal effect, we shall integrate the causal hazard
difference over $s$ to get a marginal effect. The effect will then
only be a function of $t$, and can be estimated. The average causal
hazard difference is thus:

Using Equation (\ref{individual causal}) we can then identify the ATT in
(\ref{eq:targetparameter1}) using
\begin{align}
d^*(t)=  &  \delta(t)+\gamma_{1}(t)(E^{S}(\tilde{L}_{1}^{1}(t;S))-E^{S}
(\tilde{L}_{1}^{0}(t;S)))\nonumber\\
&  +\ldots+\gamma_{k}(t)(E^{S}(\tilde{L}_{k}^{1}(t;S))-E^{S}(\tilde{L}_{k}
^{0}(t;S))).\label{first causal effect}
\end{align}

\subsection{Estimation}

\label{sec:estimation}

We will now discuss how to estimate the various components going into
$d^*(t)$.  Assume that a number of individuals, $i=1,\ldots,n$,
participate in the study.  Let $R(t)$ be the set of individuals for
which $S_{i}<t$, and who are still under observation (i.e. the event
has not happened and no censoring has occurred). Let $r(t)$ denote the
size of $R(t)$ and assume that $r(t)>0$ from time 0. For simplicity,
we assume independent censoring (see e.g. \citet{aalen08}), but in the
case of dependent censoring, one can also adjust for this using
inverse probability of censoring weighting.

First, the cumulative regression functions $\Delta(t)$ and $\Gamma
_{1}(t),\ldots,\Gamma_{k}(t),$ defined as the integrals from 0 to $t$
of $\delta(t)$ and $\gamma_{1}(t),\ldots,\gamma_{k}(t)$, are estimated
with an additive hazards model from all individuals, including those
on and those off treatment, using the \emph{actual observed} $\bm{L}$'s and
treatment status. The estimates are denoted by $\hat{\Delta}(t)$ and
$\hat{\Gamma}_{0} (t),\ldots,\hat{\Gamma}_{k}(t)$.

Then, we estimate $a_{j}(t)=E^{S}(\tilde{L}_{j}^{1}(t;S))$. When
$t>S$, then all individuals are observed to be on treatment and hence
an estimate is given as follows:
\[
\hat{a}_{j}(t)=\frac{1}{r(t)}\sum_{i\in R(t)}L_{j}(t).
\]
Note that we here make use of the consistency assumption, which means
that ``a subject's counterfactual outcome under the same treatment
regime that he actually followed is, precisely, his observed outcome''
\citep{hernan08}.

Finally, we shall estimate 
\begin{equation}
b_{j}(t)=E^{S}(\tilde{L}_{j}^{0}(t;S)).
\label{bt}
\end{equation}
This is a counterfactual quantity, and shall be estimated by
Farewell's linear increments model as described in Section
\ref{sec:flim}. The resulting estimate is denoted by
$\hat{b}_{j}(t)$. The estimated average cumulative causal effect based
on Equation (\ref{first causal effect}) is then given by

\begin{equation}
\hat{D^*}(t)=\hat{\Delta}(t)+\sum_{j=1}^{k}\int_{0}^{t}(\hat{a}_{j}(u)-\hat
{b}_{j}(u))d\hat{\Gamma}_{k}(u). \label{cum causal effect}
\end{equation}
Since we only consider treated individuals and compare with
individuals that are identical apart from counterfactually not being
on treatment, this is an estimate for a \emph{treatment effect of the
  treated}.

\section{Estimating counterfactual covariate processes when treatment
  is not started}

\label{sec:flim}

So far we have shown how to estimate the causal effect of interest
given that we can impute estimates of the confounder processes in the
counterfactual setting that treatment had not been started. We shall
view this as a missing data problem \citep{howe15}. We now give a
general description of Farewell's linear increments model and show how
this model can be used to estimate the missing counterfactual
quantities.

\subsection{Farewell's linear increments model}

The linear increments model is a dynamic model for longitudinal data,
analogous to the counting process approach for survival data. The
model was originally suggested by \citet{farewell06} and further
described and discussed in \citet{diggle07}. It was designed to
analyse, in a simple manner, missing data in longitudinal studies. A
multivariate generalisation of the model was given in
\citet{aalen10}. Applications include estimation of mean response in
studies with drop-out \citep{gunnes09} and correcting for missing data
when assessing quality of life in a randomised clinical trial
\citep{gunnes09b}. An R-package, FLIM, is available for fitting linear
increments models \citep{Hoff14}.

Let us start by imagining the complete data set (i.e. without missing
data): Let $\widetilde{K}(t)$ be an $n\times m$ matrix of multivariate
individual responses defined for a set of times $t\in\{0,\ldots,j\}$,
with $\widetilde{K}(0)=k_{0}$, where the matrix $k_{0}$ contains the
fixed starting values for the processes. The number of columns in
$\widetilde{K}(t) $ corresponds to the number of $m$ variables
measured for an individual and the number of rows corresponds to the
number of $n$ individuals.

We define the increment $\Delta\widetilde{K}(t)=\widetilde{K}(t)-\widetilde{K}%
(t-1)$ and assume, for each $t$, that $\Delta\widetilde{K}(t)$ satisfies the
model
\begin{equation}
\Delta\widetilde{K}(t)=\widetilde{K}(t-1)\beta(t)+\widetilde{\varepsilon}(t),
\label{Real model}%
\end{equation}
where $\beta(t)$ is a $m\times m$ parameter matrix and
$\widetilde{\varepsilon }(t)$ is an $n\times m$\ error matrix. The
errors are defined as zero mean martingale increments such that
$\mathrm{E}(\widetilde{\varepsilon}%
(t)\mid\mathcal{F}_{t-1})=0$, where $\mathcal{F}_{t}$ is the history
of $K(t)$ up to and including time $t$. It then follows that%

\begin{equation}
\mathrm{E}(\Delta\widetilde{K}(t)\mid\mathcal{F}_{t-1})=\widetilde{K}%
(t-1)\beta(t).\nonumber
\end{equation}

In analogy with censoring in survival analysis, missing data is common
in longitudinal data. Let $\Delta K(t)$ denote the actually observed
increments, i.e. when both components defining the increment are
observed, and set $\Delta K(t)$ equal to zero otherwise. The relation
between the true and observed responses and increments are $\Delta
K(t)=Q(t)\Delta\widetilde{K}(t)$, where $Q(t)$ is a $n\times n$
diagonal matrix defined as follows: For individual $i$, the $i$'th
diagonal element of $Q(t)$ is equal to 1 when the increment
$\Delta\widetilde{K}_{i}(t)$ is observed and 0 otherwise. Let
$\mathcal{Q}_t$ denote the history of the process $Q(t)$,
that is, up to and including time $t$.

\subsection{Causal assumptions of the linear increments model}

The key condition for proper modelling of missing longitudinal data
using the linear increments models is typically formulated trough the
assumption of discrete-time independent censoring (DTIC), as discussed
in \citet{diggle07}. This assumption, which is analogous to the
independent censoring assumption of survival analysis (see
e.g. \citet{aalen08}), places constraints on the expected values of
the increments $\Delta\widetilde{K}
(t)=\widetilde{K}(t)-\widetilde{K}(t-1)$ of the hypothetical response,
and can be formulated as follows:
\begin{equation}
  \mathrm{E}(\Delta\widetilde{K}(t)\mid\mathcal{F}_{t-1},\mathcal{Q}_{t})=\mathrm{E}(\Delta\widetilde{K}(t)\mid\mathcal{F}_{t-1}),\quad
  \mathrm{for}\text{ }\mathrm{all}\text{ }t.\nonumber
\end{equation}

The DTIC assumption has a close relationship to other no-confounders
assumptions, like sequentially missing at random \citep{seaman16}. In
order to follow the assumptions as formulated in counterfactual
analyses from causal inference, we will assume the slightly stronger
condition of sequential conditional exchangeability. This assumption
can for the modelling of missing longitudinal data be formulated as
\begin{equation}
  \Delta\widetilde{K}(t) \perp Q(t) \mid
  \mathcal{F}_{t-1},\mathcal{Q}_{t-1},\quad
  \mathrm{for}\text{ }\mathrm{all}\text{ }t \nonumber
\end{equation}
e.g. based on the definition in \citet{hernan08}.

Sequential conditional exchangeability guarantees that the observed
data will satisfy the model defined in (\ref{Real model}), so that we
can write
\begin{equation}
\Delta K(t)=Q(t)K(t-1)\beta(t)+\varepsilon(t).\nonumber
\end{equation}

\subsection{Estimation of missing covariate values}

The procedure for estimating missing covariate values is as follows:
We assume a nonparametric model over time, so that there is no assumed
connection between $\beta(t_{1})$ and $\beta(t_{2})$ for two different
times $t_{1}$ and $t_{2}$. The parameter matrices $\beta(t)$ can then
be estimated unbiasedly by the least squares approach from observed
increments, where we denote these estimates as $\hat{\beta}(t)$. The
least square estimate of $\beta(t)$ is given by
\begin{equation}
\hat{\beta}(t)=(U^{T}U)^{-1}U^{T}\Delta K(t),\nonumber
\end{equation}
where $U=Q(t)K(t-1)$.

Now, let $Q_{0}(t)$ be an indicator which is equal to 1 when the value
$\widetilde{K}(t)$ is observed. The hypothetical complete predicted data
values can be estimated iteratively by \citep{aalen10}
\begin{align}
\widetilde{K}^{\mathrm{est}}(0)  &  =k_{0},\nonumber\\
\Delta\widetilde{K}^{\mathrm{est}}(t)  &  =(1-Q_{0}(t))\widetilde{K}%
^{\mathrm{est}}(t-1)\hat{\beta}(t)+Q_{0}(t)(K(t)-\widetilde{K}^{\mathrm{est}%
}(t-1)),\quad t=1,\ldots,j\label{Iterative}\\
\widetilde{K}^{\mathrm{est}}(t)  &  =\widetilde{K}^{\mathrm{est}}%
(t-1)+\Delta\widetilde{K}^{\mathrm{est}}(t),\qquad t=1,\ldots,j\nonumber
\end{align}
given some initial values $k_{o}$. Note that when observation takes place, the
estimated value $\widetilde{K}^{\mathrm{est}}(t)$ simply equals $K(t)$. When
there is no observation, the increments are updated according to the model.

Farewell's linear increments model will now be used below for
estimating missing counterfactual values of covariate processes. For
more details on the properties of this model, see \citet{diggle07} and
\citet{aalen10}.

\subsection{Application to imputing counterfactual trajectories}
\label{sec:appimp}

The counterfactual values $b_{j}(t)$ from Equation (\ref{bt}) shall be
estimated by Farewell's linear increments model. Assuming sequential
conditional exchangeability, this model can give us a precise
prescription for how to model missing data, which is central when
estimating causal effects. Being exchangeable now means that the risk
of event among the untreated would have been the same as the risk of
event among the treated had subjects in the untreated group received
the treatment (and vice versa).  In a longitudinal setting this means
that "at every observation time $k$ and conditional on prior treatment
and covariate history, the treated and the untreated are exchangeable"
\citep{hernan08}. This is also known as a no unmeasured confounding
assumption.

\begin{figure}[ptb]
\begin{center}
\includegraphics[scale=1.1]{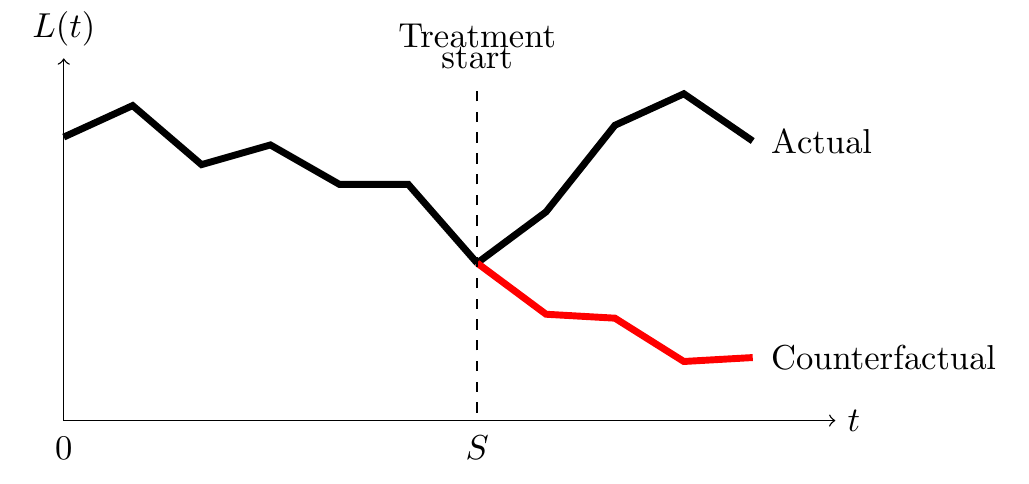}
\end{center}
\caption{ Illustrating the imputation of counterfactual covariate
  trajectories by Farewell's linear increments model.  Black line
  illustrates observed covariate trajectory and red line the
  originally unobserved, imputed counterfactual covariate trajectory
  for a hypothetical individual.}
\label{Single_counterfactual}
\end{figure}

We shall apply the linear increments model to the process
$\mathbf{L}$. We then regard the values prior to treatment start as
observed. When treatment is started the observations that would have
been made in the absence of treatment are missing. These are estimated
by the iterative procedure in Equation (\ref{Iterative}).  If
treatment is started at time $s$ we thus estimate the quantity $\tilde
{L}_{j}^{0}(t;s)$\ for each $j$. Taking an average of these at time
$t$ over all individuals in $R(t)$ gives an estimate of
$b_{j}(t)$. Notice from the definition of $R(t)$ in Section
\ref{sec:estimation} that estimation of counterfactual trajectories
will only take place until the individual is censored or experiences
an event in the observed data. An illustration of the imputation idea
is given in Figure \ref{Single_counterfactual}.

An example showing how to impute counterfactual covariate values using
the linear increments models has recently been included in the R
package FLIM \citep{Hoff14}.

\section{A mediation point of view}
\label{sec:mediation}

Mediation can also be studied for causal effects defined in an ATT
setting, see e.g. \citet{vansteelandt12natural}. We shall show that
the formula for $d^*(t)$ in Equation (\ref{first causal effect}), can
be interpreted as the sum of a direct effect on the treated
$\delta(t)$ and of an indirect effect transmitted through the
covariates. This can be made precise as follows:

The natural direct effect is the effect that would be observed if the
mediator under treatment was manipulated to be equal to the natural
value of the mediator under no treatment. However, in our survival
setting it is possible that the individual might survive in one
counterfactual world and not in the other one, therefore manipulation
at the individual level may not make sense. Instead we can use the
concept of randomized interventional analogues of natural direct and
indirect effects from \citet{vanderweele15}, see also standardized
direct effects in \citet{didelez06ria}.

Consider individuals that have started treatment. According to
\citet[Section~5.4.1]{vanderweele15}; ``we will consider what would
have happened if we fixed their mediator to a level that is drawn
randomly from the subpopulation that is unexposed. Thus instead of
using the individual’s particular value for the mediator in the
absence of exposure, we use the distribution of the mediator amongst
all the unexposed''. Our procedure follows the same idea in a slightly
more general way, namely using the distribution of the mediator values
among those not on treatment to estimate values of the mediator for
those on treatment, where they not treated. This is precisely the
calculation that is done by Farewell's linear increments model.

For the purpose of estimating average effects we can formulate this
for expected values. Assume that the mediator is the set of all
time-dependent covariates; hence, we assume that $L_{i}^{1}(t;s)$ is
manipulated to be equal to $L_{i}^{0}(t;s)$ for all $i$. The covariate
part in Equation (\ref{first causal effect}) would then be equal to 0,
and so we would have
\begin{equation}
d^*(t)_{dir}=\delta(t)
\label{eq:direff}
\end{equation}
as the average direct effect defined in the above sense. The average
indirect effect is defined as the difference between the total causal
effect and the direct effect; that is
\begin{align}
d^*(t)_{indir}  &  =\gamma_{1}(t)(E^{S}(\tilde{L}_{1}^{1}(t;S))-E^{S}(\tilde
{L}_{1}^{0}(t;S)))\nonumber\\
&  +\ldots+\gamma_{k}(t)(E^{S}(\tilde{L}_{k}^{1}(t;S))-E^{S}(\tilde{L}_{k}
^{0}(t;S))). 
\label{eq:indireff}
\end{align}

\begin{figure}[ptb]
\begin{center}
\includegraphics[width=0.65\textwidth]{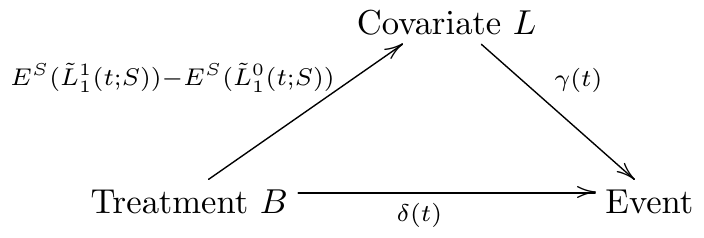}
\end{center}
\caption{Illustration of the mediation model for the causal effect of
  treatment on a time to event outcome from Equation (\ref{eq:direff})
  and (\ref{eq:indireff}) in the simple case of $k=1$.}
\label{mediation}
\end{figure}

A diagram illustrating a mediation model for $k=1$ is shown in Figure
\ref{mediation}. The figure indicates that the causal effect may be
seen as a sum of a direct effect, and an indirect effect passing
through the covariate; the latter being the product of the
coefficients on the two arrows.

Note that we have here a well defined composition valid at any time
$t$. Based on this we can estimate cumulative direct and
indirect effects from Equation (\ref{cum causal effect}), that is:
\begin{equation}
\hat{D^*}(t)_{dir}=\hat{\Delta}(t),\qquad\hat{D^*}(t)_{indir}=\sum_{j=1}^{k}
\int_{0}^{t}(\hat{a}_{j}(u)-\hat{b}_{j}(u))d\hat{\Gamma}_{k}(u).
\end{equation}

\section{A shortcut through regression}
\label{sec:shortcut}

A simple regression analysis can give us an approximate estimate of
the causal effect in Equation (\ref{cum causal effect}). To do such an
analysis one only needs to manipulate the values of the time-varying
covariates in the original dataset, by giving individuals on treatment
the imputed covariate values which they would have had, if they were,
counterfactually, untreated. The manipulated dataset can then be
analysed in a standard regression, adjusting for treatment, baseline
covariates and the partly manipulated time-varying covariates. The
regression function for the treatment covariate would then be the
approximate estimate of the causal treatment effect.

Why should this work? The idea is that if those on treatment are given
counterfactual covariates corresponding to no treatment, then the
difference in outcome between treated and untreated must be due to the
treatment. The same idea has worked well in the two-stage approach of
\citet{kennedy2010} and \citet{taylor2014}. The setting of these
papers are different but the basic concept is the same.

\subsection{The additive hazards model}
\label{seq:addshortcut}

The validity of the shortcut is best shown for additive models. We
shall show later, in simulations, that the results are very close to
those found by the direct approach in Section \ref{sec:additive
  model}. However, we do not get exact theoretical results as we did
in Section \ref{sec:additive model}.

To put it formally: For an individual starting on treatment at time
$s$, the hazard rate of an event from Equation (\ref{eq:h1ts}) can be
rewritten as follows $\mathrm{for\;}t\geq s$:

\begin{align*}
h^{1}(t;s)  &  =\delta(t)+\gamma_{0}(t)
c+\gamma_{1}(t)\tilde{L}_{1}^{1}(t;s)+\ldots+\gamma_{k}(t)\tilde{L}_{k}
^{1}(t;s)\\
&  =d(t;s)+\gamma_{0}(t)
c+\gamma_{1}(t)\tilde{L}_{1}^{0}(t;s)+\ldots+\gamma_{k}(t)\tilde{L}_{k}
^{0}(t;s)
\end{align*}
For individuals not on treatment, that is, $t<s$, the hazard rate is
given by

\[
h(t;0,\mathbf{L})=\gamma_{0}(t)
c+\gamma_{1}(t)L_{1}(t)+\ldots+\gamma_{k}(t)L_{k}(t).
\]
This is consistent with the following additive hazards regression model
applied to all individuals at risk:
\[
\mu(t)=\gamma_{0}(t)
c+d(t,s)B(t)+\gamma_{1}(t)L_{1}^{\ast}(t,s)+\ldots+\gamma_{k}(t)L_{k}^{\ast
}(t,s),
\]
where $B(t)$ is the treatment indicator as defined earlier, and
$L_{i}^{\ast}(t)$ equals $\tilde{L}_{j}^{0}(t;s)$ if $B(t)=1$ and
$L_{i}(t)$ if $B(t)=0$. Note that for those on treatment we put in
counterfactual values of the covariates to mimic what they would have
were they not on treatment; this is the crux of the procedure. The
causal effect for an individual starting treatment at time $s$ is seen
to equal the coefficient of the variable $B(t)$.\ Note that this
coefficient is dependent on the value of $S$, and is hence random
(i.e. varying between individuals). If we knew the values of
$\tilde{L}_{j}^{0}(t;s)$ and could run this regression we would expect
that the coefficient corresponding to the covariate $B(t)$ would be an
average of the possible values, and thus a reasonable estimate of
$d^*(t)$. Since we estimate the cumulative coefficient in the additive
regression model, we actually estimate $D^*(t)$, denoting this
estimate $\tilde{D^*}(t)$. This is an approximate procedure since the
coefficient of $B(t)$ varies with the individual treatment starting
time $S$, thus we have a varying coefficient regression model.
However, simulations show that $\tilde{D^*}(t)$ is very close to
$\hat{D^*}(t)$ as might be expected. In fact, the simulations in
Section \ref{sec:simulation} indicate that the two approaches estimate
almost exactly the same thing.

The procedure depends on estimating the $\tilde{L}_{j}^{0}(t;s)$,
which shall be done as previously, using Farewell's linear increments
model. Since we have a precise argument for $\hat{D^*}(t)$ and a less
precise one for $\tilde{D^*}(t)$, the first one might be seen as more
reliable. However, $\tilde{D^*}(t)$ is slightly easier to compute and
fits with ideas already in the literature \citep{kennedy2010,
  taylor2014}.

\subsection{The Cox model}

Since the Cox model is the most common one in survival analysis, it is
natural to ask whether the present approach could be applied to this
model as well.  The formal mathematical arguments in Section
\ref{sec:additive model} and the causal effect given in Equation
(\ref{cum causal effect}) does not work in this case; however, the
intuitive argument given in Section \ref{sec:shortcut} makes sense for
the Cox model as well. Hence, one would expect that the shortcut
method might work for the Cox model. The hazard rate would then be:%

\[
\mu(t)=\exp(\gamma_{0}(t)
c+d(t,s)B(t)+\gamma_{1}L_{1}^{\ast}(t,s)+\ldots+\gamma_{k}L_{k}^{\ast}(t,s))
\]

Again, we have a random coefficient model. What will be estimated by a
Cox model is some kind of average of the quantity
$d(t,s)$. Preliminary simulation in Section \ref{sec:simulation cox}
seems to indicate that this might give sensible results, but further
work on this issue is required.

\section{Application to data from the Swiss HIV Cohort Study}

\label{sec:shcs}

Let us now consider data from \citet{shcs10}, studying the effect of
antiretroviral treatment (HAART) on time to AIDS or death. The dataset
we analyse consists of 2161 HIV infected individuals, with baseline at
the time of the first follow-up after January 1996. The data is
organised in monthly intervals, with time-varying variables describing
the treatment received, CD4 cell count, viral load (HIV-1 RNA),
haemoglobin levels and other relevant clinical variables. Scheduled
clinical follow-up with protocol defined laboratory tests takes place every
sixth month and on average one additional intermediate routine
laboratory test is also recorded. In months with no new observations,
the last observation is carried forward. Baseline variables include
sex, year at birth, registration date and transmission category. The same dataset
has been analysed before in \citet{sterne05, gran10, roysland11}.

When estimating the effect of antiretroviral treatment on time to AIDS
or death, time-dependent prognostic factors such as CD4, viral load
and haemoglobin levels are typically time-dependent
confounders. Time-dependent confounding can be adjusted for using
marginal structural models (see \citet{sterne05}) or the sequential
Cox approach (see \citet{gran10}). We shall here apply our new method
for analysing the ATT.

\begin{figure}[ptb]
\centering
\includegraphics[height=5.4in]{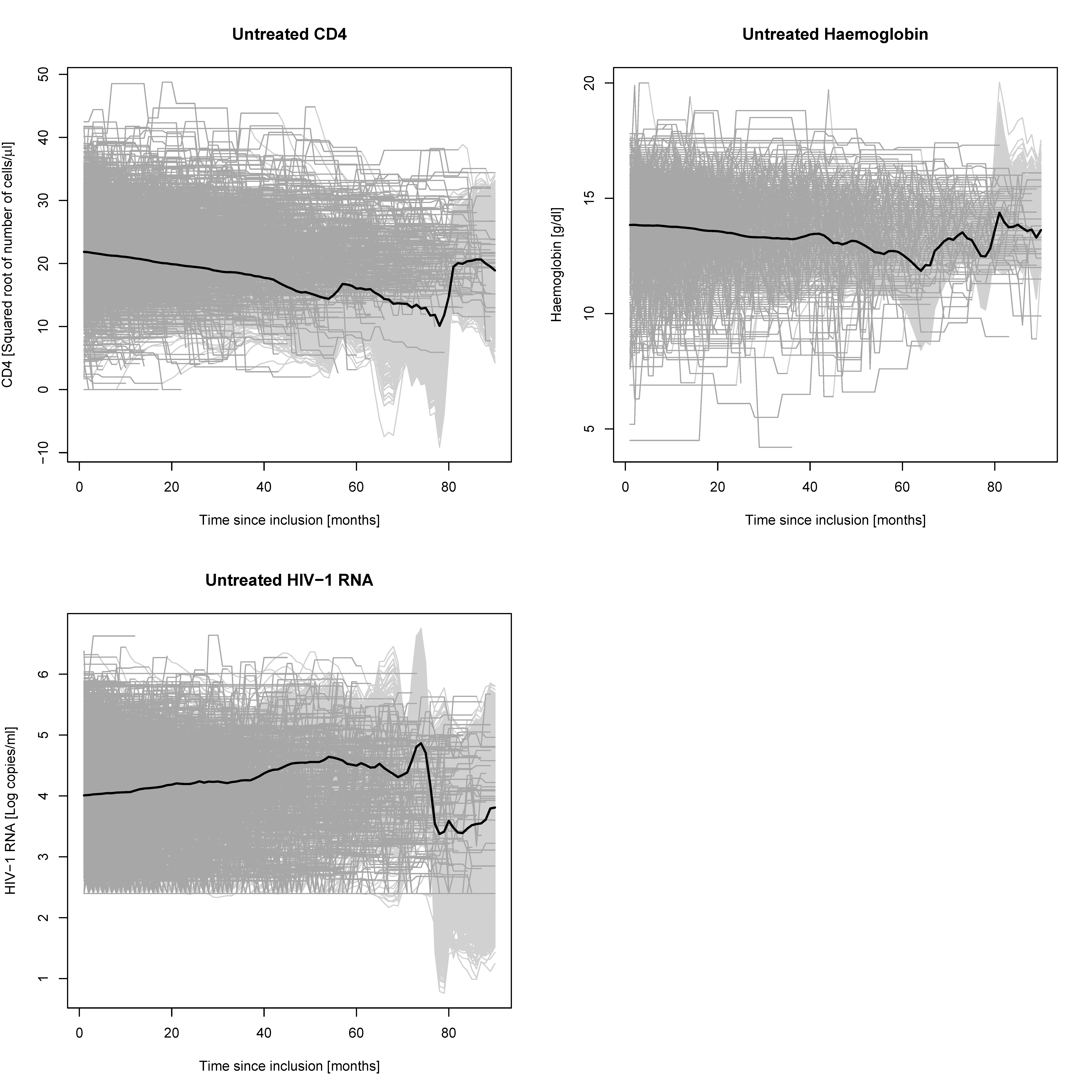}\caption{Observed (dark grey),
estimated (light grey) and mean (black) CD4, viral load (HIV-1 RNA) and
haemoglobin trajectories for untreated individuals.}
\label{fig1}
\end{figure}

Let us consider the three time-dependent variables CD4 count, viral
load and haemoglobin level. For all individuals in the dataset that
started treatment, we estimate their counterfactual covariate
trajectories had they stayed untreated, using the iterative procedure
described in Section \ref{sec:flim}.  The counterfactual covariate
values are estimated from the time treatment started and until the
individuals are censored or experience an event in the observed
data. When estimating counterfactual covariates, we adjust for
baseline variables sex, age at baseline, year at baseline and
transmission category, together with the three time-dependent
variables. The observed and estimated covariate trajectories for all
untreated individuals are shown in Figure \ref{fig1}.

Figure \ref{fig1} shows that over time untreated individuals
experience a decrease in CD4 and haemoglobin levels, and an increase
in viral load. We see that the estimated (light grey) covariate
trajectories typically depict individuals who are worse off, because
they are modelled counterfactual trajectories, that is; trajectories
for individuals who have started treatment in the observed data. This
is best seen in the CD4 cell count trajectories because CD4 cell count
is the most important predictor of treatment initiation. Note that
model uncertainty gets bigger with time, as the size of the risk set
decreases. This is seen by the increasing fluctuations of estimated
CD4 counts with time. Because of this uncertainty the plots are
truncated at 80 months, even though the last observation was 92 months
after the start of follow-up.

\begin{figure}[ptb]
\centering
\includegraphics[height=5.6in]{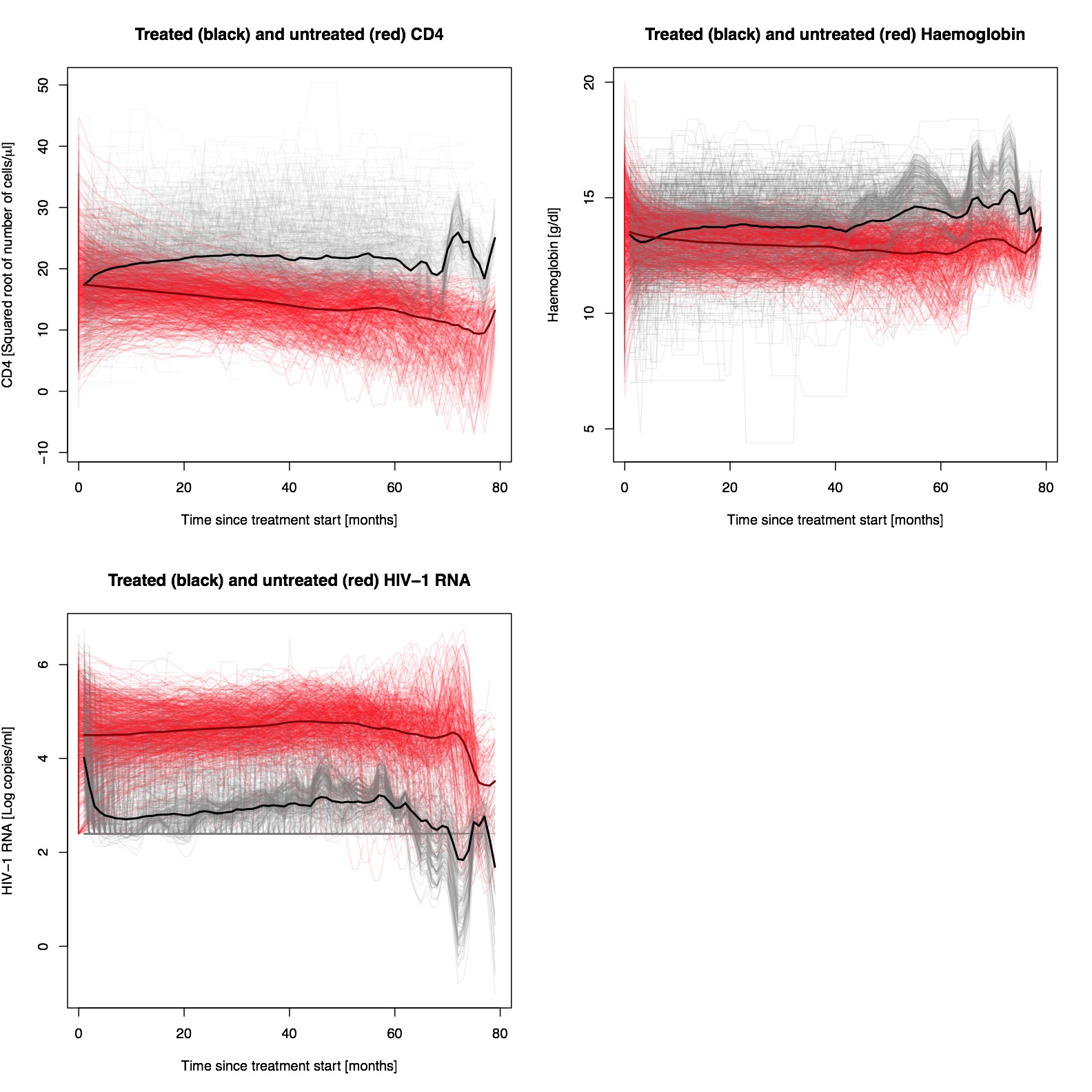}\caption{Observed, estimated
  and mean CD4, viral load (HIV-1 RNA) and haemoglobin trajectories
  for the two counterfactual regimes of treatment (black) and no
  treatment (red).}
\label{fig2}%
\end{figure}

The effect of treatment on the time-dependent covariates themselves
can be studied graphically, by changing the time-scale to time since
start of treatment and plotting the two counterfactual regimes of
treated and untreated together (Figure \ref{fig2}). Here, the
untreated group (the red lines) represent the unobserved
counterfactual covariate trajectories the treated individuals would
have had if they were not treated, which are all imputed values. For
the treated group (black and grey lines), imputations are only needed
when patients are censored. The rest of their trajectories are
observed under treatment, as no individual, by definition, goes from
being treated to untreated.

In Figure \ref{fig2} we clearly see the negative consequences for
untreated patients; lower CD4 counts and haemoglobin levels and higher
viral load.  Treated individuals, on the other hand, experience a rise
in CD4 cell counts and a rather rapid decline in viral load.

Let us now return to the main objective of adjusting for
time-dependent confounding. We want to estimate the treatment effect
on the treated for antiretroviral treatment on time to AIDS or death,
using the method in Section \ref{sec:additive model}. In practice, we
impute values for all counterfactual time-varying covariates under the
scenario of no treatment, wherever these variables were not
observed. The only time-varying variable we do not alter is the
treatment indicator, representing the exposure we want to estimate,
that is; the effect of treatment on time to AIDS or death.

We use the shortcut approach in Section \ref{sec:shortcut} since it
was demonstrated in Section \ref{sec:simulation} that it gives almost
exactly the same result as the causal effect formula (\ref{cum causal
  effect}). The analysis was carried out fitting an additive hazards
model to a pseudo dataset where all time-varying covariates for
individuals on treatment were imputed using the linear increments
model, adjusting for possible non-random dropout using inverse
probability of censoring weights. The included covariates were HAART,
sex, transmission category, baseline CD4 (sqrt cells per $\mu$L), baseline RNA
(log10 copies per mL), baseline haemoglobin (g per mL), baseline CDC B
event, previously experienced CDC B event at baseline, and current
CD4, RNA and haemoglobin. The data is grouped into monthly intervals
and baseline is set to the time of the first follow up visit after
January 1996. For the marginal structural additive hazards model, the
same covariates were used, except that the time-varying covariates
were replaced by stabilised inverse probability of treatment
weights. The applied censoring and treatment weights are identical to the weights used in
\citet{sterne05}, which analysed the same dataset.

\begin{figure}[ptb]
\centering
\includegraphics[height=2.4in]{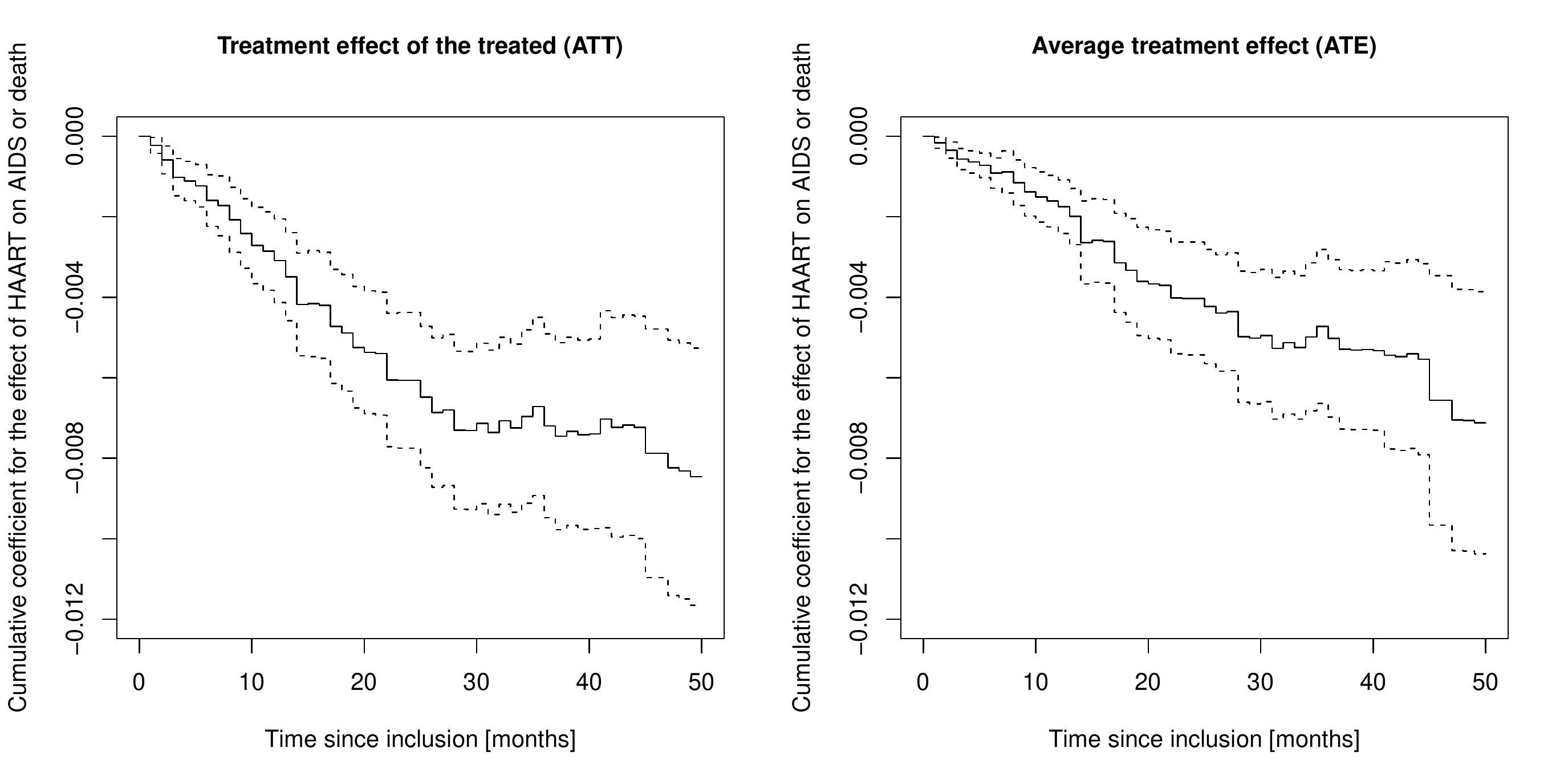}\caption{Analysis of the
  effect of HAART on AIDS or death in the Swiss HIV Cohort Study
  comparing estimates of the treatment effect on the treated (left
  panel) with average treatment effects (right panel). Both analyses
  are based on additive hazards regression, the first imputing
  counterfactual covariate trajectories from a linear increments model
  and secondly using an inverse probability weighted marginal structural
  model.}
\label{fig3}
\end{figure}

Figure \ref{fig3} (left panel) shows the cumulative treatment effect
on the treated for HAART versus no treatment, found using the approach
in Section \ref{seq:addshortcut}. Treatment has a constant protective
effect (a negative contribution to the hazard) for the first 30
months, before the effect decreases. The plot is truncated at 50
months to give a more detailed picture of the initial phase. As a
comparison, in the second panel of Figure \ref{fig3}, we include the
cumulative coefficient for HAART in a inverse probability weighted
marginal structural Aalen additive model. This is merely a weighted
additive hazards model using the same inverse probability of treatment
and censoring weights as used in the marginal structural Cox
proportional hazards model in \citet{sterne05}. We see from Figure
\ref{fig3} that the estimated cumulative effects of treatment are
somewhat smaller when using the marginal structural model. This is
reasonable, since one would expect those actually treated to have a
greater potential treatment effect on the average than those not
actually treated.

Tests of effect (i.e. testing for slope) may be carried out for the
plots in Figure \ref{fig3}, giving $p=0.0003$ for the left panel and
$p=0.002$ for the right panel.

Note that robust confidence intervals are used for the curves in
Figure \ref{fig3}. The intervals in the left panel may be slightly too
narrow, as the uncertainty in the imputed values are not brought into
the calculations of these intervals. An alternative is to calculate
confidence intervals by bootstrapping.

All analyses were done in R version 3.0.0, using the \texttt{aareg}
function from the package \texttt{survival}.

\section{Simulation study}

\label{sec:simulation}

Let us now analyse simulated data to compare the methods and study
their behaviour. The methods we want to compare are the procedures
based on imputation of counterfactual covariate trajectories for
estimating ATT, inverse probability weighted marginal structural
models for estimating ATE and naive methods that do not properly
adjust for time-dependent confounding.

To generate data with time-dependent confounding we simulate event
times from an additive hazards model.  We pretend to have a cohort of
individuals under risk of experiencing some event of interest.  Individuals
are routinely controlled to have a covariate value measured.  The
probability of starting treatment on a specific time point will depend
on an individual's current covariate value, and in return, starting
treatment affects the future covariate process. Both treatment status
and the covariate value affect the probability of having an event.

\subsection{Data generation}

Imagine a cohort of $n$ individuals under risk of having an event
$E(t)$, which can be postponed or prevented by treatment; everyone in
the cohort starts out untreated.  Each individual has a time-varying
covariate $L(t)$ repeatedly measured on a set of pre-specified time
points $t = 0,1,...,11$, given that they have not yet experienced the
event.  To generate initial values of $L(t)$ for each subject, we took
the square root of $n$ numbers uniformly distributed on the interval
$[25, 1000]$. The range of the interval and the square root
transformation were employed to obtain a variable somewhat comparable
to CD4 cell count. Untreated, $L(t)$ will decrease steadily, while
when receiving treatment, $L(t)$ increase over time. Individuals who
start treatment are not allowed to go back to being untreated. The
hazard of having an event within an interval $[t, t+1)$ is affected by
treatment status $B(t)$ and by covariate value $L(t)$. Everyone enter
the study at $t=0$, and those who have not experienced the event by
$t=11$ are censored.

We will consider three different treatment regimes, which are
summarised in Figure 1 in the online appendix accompanying this paper.
In the first, people with low covariate values have a high probability
of starting treatment. The second regime is close to randomised
treatment, meaning there are only small differences between treated
and untreated individuals in terms of covariate values.  Finally, the
last regime is the opposite of the first, i.e, individuals with higher
covariate values are prioritised for treatment. The online appendix
also contains a pseudo code algorithm for simulating data under these
three treatment regimes and the full corresponding R code.  We ran the
simulations 250 times, and in each run we generated a new dataset that
was analysed with the methods described in the following section.

\subsection{Analysing simulated data}

The simulated data is analysed with the procedures for estimating the
ATT as described in Section \ref{sec:estimation} and Section
\ref{sec:shortcut}.  For comparison, we also estimate the ATE from a
marginal structural additive hazards model and treatment effects from
two naive approaches. Implementing the marginal structural model
involves estimating weights based on the inverse probabilities of
starting treatment, which are then used in a weighted regression. The
marginal structural model does correct for time-dependent confounding,
but unlike our methods, the estimated effect of treatment is
equivalent to comparing the effect of everyone being treated to no one
being treated, as illustrated in Figure \ref{ATT}. The naive methods
are additive regression analyses that do not properly correct for
time-dependent confounding.  In the first naive method we control for
both treatment and covariate, and in the second we only control for
treatment.

All of the methods are different implementations of Aalen's additive
hazards regression models, and we are interested in estimating the
effect of treatment on survival. The estimated effects are displayed
as curves showing the differences in cumulative hazards between the
treated and untreated. 

For the simulations, there are no single target parameter we can
compare against to assess performance and check that methods are
valid.  Instead, we include an analysis of a simulated \textit{full
  counterfactual dataset}.  This dataset is constructed by combining
data coming from the algorithm shown above where treatment is offered,
with data from the same algorithm, but without offering
treatment. This means that we end up with a dataset where those who
originally received treatment are also included with their untreated
histories.  Analysing such a dataset with a univariable regression,
controlling only for treatment, yields an estimate of the average
treatment effect on the treated (ATT). In the simulation results, we
will call this \textit{treatment effect on the treated: simulated}.

Finally, we analyse the effect of treatment when starting treatment is
randomised. This corresponds to finding the average treatment effect
(ATE). More specifically, we implement a univariate regression,
controlling only for treatment, to a dataset generated the same way as
earlier, but with a flat probability set to 0.07 for starting
treatment on all time points, for all individuals regardless of
covariate values. The value 0.07 was chosen so that the random
treatment setting was reasonably comparable to the other treatment
regimes in terms of proportions of the cohort starting treatment at
some point during follow-up.

\subsection{Simulation results for the additive model}

\begin{figure}[ptb]
\begin{center}
\includegraphics[width = 0.9\textwidth]{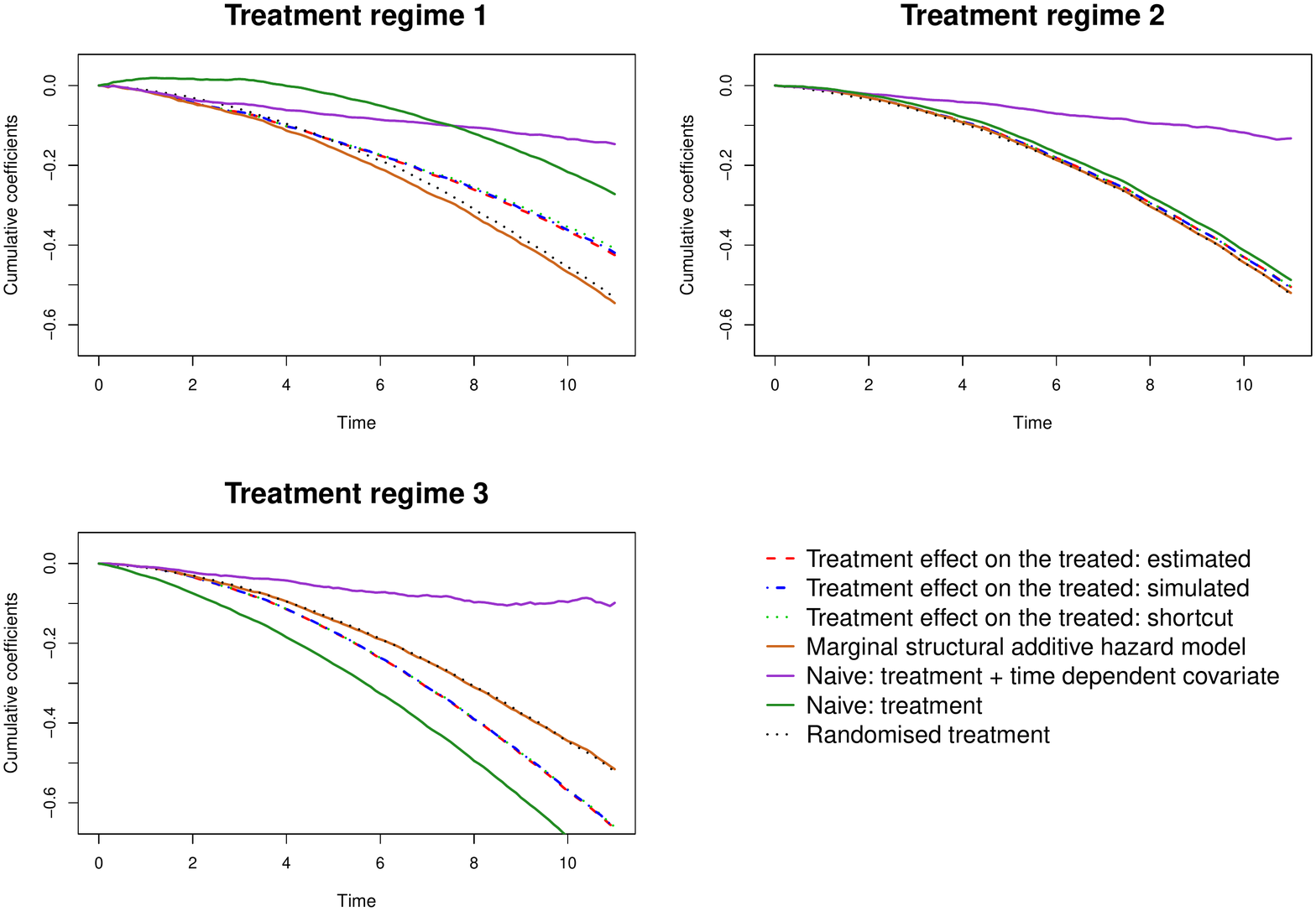}
\end{center}
\caption{Average cumulative coefficients for the treatment effect for the
three treatment regimes in the simulation study.}
\label{sim1}
\end{figure}

Figure \ref{sim1} shows the mean of the cumulative hazard differences
for being on treatment compared to not being on treatment. Both our
proposed methods for estimating ATT in Section \ref{sec:estimation}
and Section \ref{sec:shortcut}, the red curve and the dotted green
curve, show similar estimates as the simulated target curve (blue)
across all treatment regimes.

Our methods estimate the treatment effect on the treated. This effect
varies between different treatment regimes. The marginal structural
additive hazards model on the other hand, gives more or less the same
curve estimates across the different regimes -- estimates that are
similar to the estimated treatment effect in a randomised study, the
black dotted curve. The naive method where we control for both
treatment and covariate value is estimating the direct effect of
treatment, which is constant across all three simulations (the purple
curves).  The second naive method, green full-drawn line, where we
only control for treatment, behaves very different from the advanced
methods. In all but the second regime it fails to estimate a valid
parameter, as it does not properly correct for the time-dependent
confounding that is strong in regime 1 and regime 3.

The simulations show how our proposed methods estimate treatment
effects that are dependent on who actually get treated.  Given that we
can impute counterfactual covariate trajectories, the procedures
succeed in estimating the target curve that comes from analysing the
full counterfactual dataset.

\subsection{Simulation results for the Cox model}

\label{sec:simulation cox}

\begin{table}[th]
\centering
\begin{tabular}
[c]{rrrr}\hline\hline
\textbf{Treatment regime:} & \textbf{1} & \textbf{2} & \textbf{3}%
\\\hline\hline
Treatment effect on the treated: simulated & 0.77 & 0.74 & 0.69\\
Treatment effect on the treated: shortcut & 0.79 & 0.75 & 0.68\\
Marginal structural model & 0.75 & 0.74 & 0.75\\
Naive: treatment + time dependent covariate & 0.93 & 0.93 & 0.91\\
Naive: treatment & 0.84 & 0.76 & 0.64\\
Randomised treatment & 0.75 & 0.75 & 0.75\\\hline
\end{tabular}
\caption{Cox model average hazard ratios for the effect of
  treatment.}
\label{Cox table}
\end{table}

Simulating data from Cox proportional hazards models with
time-dependent confounding can be a very challenging task
\citep{havercroft12}, so instead we use the data already simulated for
the additive model where, in contrary, simulation is very simple. This
makes sense as it is quite clear that the Cox model, as well as other
models, typically are used for data where the assumptions are not
exactly fulfilled. In fact, one of the useful aspects of the Cox model
is its robustness towards many types of model deviations.

The results from applying a Cox based analysis to the data simulated
in Section \ref{sec:simulation} are presented in Table \ref{Cox
  table}. We give all the same analyses as for the additive model
apart from that based on Equation (\ref{cum causal effect}). Hence,
the various settings are the same as previously and can be directly
compared with the results for the additive model. First, one can see
(in lines 1 and 2) that estimating the hazard ratio for the ATT using
the shortcut approach gives essentially the same results as when
simulating the same scenario. This indicates that the estimation
method gives sensible results. In line 3 one sees that the marginal
structural Cox model result in a hazard ratio that is independent of
the treatment regime, as it should be, and that this fits with the
randomised treatment (last line). The ATT analysis gives weaker
effects than the ATE from the marginal structural model for regime 1,
the same effect for regime 2, and stronger effects for regime 3. This
again fits with the results for the additive model, and is as
expected. The naive analyses also have the same type of deviations as
seen for the additive model.

These are merely preliminary results, and obviously the issue of the
validity of our approach for the Cox model needs further
clarification.

\section{Discussion}

\label{sec:discussion}

We have presented a new approach for estimating the ATT under
time-dependent confounding. The ATT is a useful alternative to the
ATE, and the two causal estimands answer two different questions. The
latter answers the question \emph{how effective is the treatment if
  treatment was randomised over time?'}, while the ATT answer the
slightly different question of \emph{how effective was treatment for
  those who received it at the time they did?} This is a subtle, but
important difference, and it is not given that one should
automatically use one method or the other when analysing data with
time-dependent confounding -- it depends on what questions one asks.

The ATT is a relevant parameter, both in a clinical and
epidemiological setting, and quantifies the average effect of
treatment as it was actually given. In other words; it quantifies the
effect of the current treatment policy and not hypothetical ones. The
ATT is often claimed to be a more relevant and preferred effect
measure than the ATE in settings with selection among treated patients
\citep{li14}. This is obviously the case in situations with
time-dependent confounding, such as in the application to the Swiss
HIV Cohort Study. The ATT for HIV treatment has been the target
parameter in both early \citep{robins92, hernan05g} and recent papers
\citep{wallace16} using g-estimation. In settings were both the ATE
and ATT can be identified, such as in our application, estimating both
and comparing them will also add to the overall knowledge of the
treatment effect. A difference between the two effect measures gives
an indication on the effect of the current treatment policy.

Using the additive model gives an explicit derivation of the ATT in
terms of counterfactual covariate processes. The positivity assumption
is not as strict as when estimating the ATE, e.g. using inverse
probability weighted marginal structural models, because there is no
requirement that each individual should have a positive treatment
probability. The no unmeasured confounding assumption is however still
central. This also goes for the linear increments model when
estimating counterfactual covariate processes, which we discussed
earlier in terms of assuming conditional exchangeability. Another
important assumption is the validity of the linear increments model
itself. See \citet{vanderweele13sens} for more on sensitivity analysis
of unmeasured confounding in additive and proportional hazards models
and \citet{brumback04sens} for more on sensitivity analysis with
time-dependent confounding in HIV cohort studies.

Our approach for estimating the ATT also has the benefit of giving a
detailed picture of how treatment works on the time-dependent
variables themselves, and not only on the outcome. This of course
comes at the cost of the assumptions for using the linear increments
model, but in the application to data from the Swiss HIV Cohort Study
we found that the this model provided stable and plausible estimates
of counterfactual covariate trajectories. We have then shown that
these estimates can be used to estimate the ATT in the presence of
time-dependent confounding in survival data, by imputation in the
additive hazards model. This can be seen as a front-door type of
approach \citep{pearl09}, where causal effects are identified by
estimating mechanisms, and our method may therefore also give
additional insight into the dynamics of how treatment works.

Note that in our use of the linear increments model, the
counterfactual covariate trajectories are only estimated until an
individual experiences an event or is censored, as it can be seen in
Figure \ref{fig1}. Covariate values beyond such events are not needed
in order to fit our model for the overall treatment effect. However,
for the sake of studying the effects of treatment initiation on the
time-varying covariates themselves (and not on the main outcome) later
covariate values might be useful. In that case, the linear increment
model can be used to estimate covariate values after the time of
censoring (or after the time of an event in the more conceptual case
of immortal cohorts). See e.g. \citet{diggle07} for a further
discussion of such topics.

Note also that the linear increments model is only one of many
possible models that could be used for modelling of counterfactual
covariate trajectories. Our approach could equally be used with other
methods. Models that describe the progression of a specific disease
using differential equations are another option, such as the HIV model
in \citet{prague13}. Different models, including a version of
Farewell's linear increments model, have also been compared in other
papers \citep{prague15}.

Compared to other methods that estimate the ATT under time-dependent
confounding on a time-to-event outcome our method is a two step
approach, where each step has the benefit of being easy to implement
using two simple existing statistical software packages. The outcome
model also have the benefit of being a hazard regression model in the
traditional sense, which typically is not the case in the g-estimation
approach. We believe that there are advantages in considering several
approaches for handling the thorny issue of time-dependent
confounding, and that the procedure described in this paper serve as a
valuable addition.

\section{Acknowledgement} 

Jon Michael Gran, Rune Hoff and Odd O. Aalen were partly funded by the
Research Council of Norway, project numbers 191460 and 218368. Kjetil
Røysland was partly supported by the Norwegian Cancer Society
contract/grant number 2197685.

\bibliography{draft}

\end{document}